\documentclass[12pt]{article}

\newcommand{\be}{\begin{equation}}
\newcommand{\ee}{\end{equation}}
\newcommand{\bi}[1]{\vspace{-3mm} \bibitem{#1}}

\usepackage{epsfig,amsmath,amssymb,graphics,graphicx} 

\begin{document}

\begin{center}
{\Large \bf Fractional Stability}
\vskip 5 mm

{\large \bf Vasily E. Tarasov }\\

\vskip 3mm

{\it Skobeltsyn Institute of Nuclear Physics, \\
Moscow State University, Moscow 119991, Russia } \\
{E-mail: tarasov@theory.sinp.msu.ru}\\
\end{center}

\begin{abstract}
A fractional generalization of variations 
is used to define a stability of non-integer order. 
Fractional variational derivatives are suggested to describe 
the properties of dynamical systems at fractional perturbations. 
We formulate stability with respect to motion changes 
at fractional changes of variables.
Note that dynamical systems, which are unstable "in sense of Lyapunov", 
can be stable with respect to fractional variations.
\end{abstract}

%%%%%%%%%%%%%%%%%%%%%%%%%%%%%%%%%%%%%%%%%%%%%%%%%%%%%%%%%%%%%%%%%%%%%%%%%%

The theory of integrals and derivatives of non-integer order goes back 
to Leibniz, Liouville, Riemann, Grunwald, and Letnikov \cite{SKM,KST}. 
Fractional analysis has found many
applications in recent studies in mechanics and physics.
The interest in fractional integrals and derivatives 
has been growing continuously 
during the last few years because of numerous applications \cite{Book2011}. 
In a short period of time the list of applications has been expanding. 
For example, it includes the chaotic dynamics \cite{Zaslavsky1,Zaslavsky2},
material sciences \cite{Hilfer,C2,Nig1}, 
mechanics of fractal and complex media \cite{Mainardi,Media},
quantum mechanics \cite{Laskin,Naber}, 
physical kinetics \cite{Zaslavsky1,Zaslavsky7,SZ,ZE,FSM},
long-range dissipation \cite{GM,TZ2}, 
non-Hamiltonian mechanics \cite{FracHam,FracVar},
long-range interaction \cite{Lask,TZ3,Map}. 

In this preprint, we use the fractional generalization of 
variation derivatives that are suggested in \cite{FracVar,Book2011}. 
Fractional integrals and derivatives are used for
stability problems (see for example \cite{S1,S2,S3,S4,S5}). 
We consider the properties of dynamical systems
with respect to fractional variations \cite{FracVar}.
We formulate stability with respect to motion changes 
at fractional changes of variables (see \cite{FracStab} pages 294-296).
Note that dynamical systems, which are unstable "in sense of Lyapunov" 
\cite{Malkin,Demidovich,Tchetaev}, 
can be stable with respect to fractional variations. 

%%%%%%%%%%%%%%%%%%%%%%%%%%%%%%%%%%%%%%%%%%%%%%%%%%%%%%%%%%%%%%%%%%%%%%%%

Let us consider a dynamical system that is desribed
by the differential equations
\be \label{3-3-1a}
D^1_t x_i = F_i({\bf x}) ,\quad i=1,...,n.
\ee
where $x_1,...,x_n$ are real variables that define
the state of the system.

We can consider variations $\delta x_i$ of the variables $x_i$.
The unperturbed motion is satisfied to
zero value of the variations, $\delta x_i=0$.
The variations  $\delta x_i$ describe
as function $f({\bf x})$ at arguments $x_i$ change varies.

Let us consider the case $n=1$.
The first variation describes a function change
at the first power of argument change:
\be
\delta f(x) = \delta x \, D^1_x f(x) .
\ee
The second variation describes a function change
at the second power of argument change:
\be
\delta^2 f(x) = (\delta x)^2 \, D^{2}_x f(x) .
\ee
The variation $\delta^{n}$ of integer order $n$ is defined
by the derivative of integer order $D^{n}_x f(x)$, such that
\[ \delta^n f(x) = (\delta x)^n \, D^{n}_x f(x) . \]

We can define \cite{FracVar,Book2011} a variation of fractional order
$m-1 < \alpha \leqslant m$ by the equation
\be \label{3-fv}
\delta^{\alpha} f (x) = [\delta x]^{\alpha} \, _0^CD^{\alpha}_{x} f (x) ,
\ee
where
\[ [\delta x]^{\alpha} = (sgn(x))^{m} |\delta x|^{\alpha} , \]
and $_0^CD^{\alpha}_{x}$ is the Caputo fractional derivative \cite{SKM,KST}
with respect to $x$.
The fractional variation of order $\alpha$ describes the function
change at change of the fractional power of argument.
The variation of fractional order is defined by the derivative
of fractional order.

Let us obtain the equations for fractional variations
$\delta^{\alpha} x_i$.
We consider the fractional variation of equation (\ref{3-3-1a}) in the form:
\be \label{3-fvar1}
\delta^{\alpha} D^1_t x_i = \delta^{\alpha} F_i({\bf x}) ,\quad i=1,...,n.
\ee
Using the definition of fractional variation (\ref{3-fv}), we have
\be \label{3-Sta1}
\delta^{\alpha} F_i({\bf x})=
\, _{a_j}^CD^{\alpha}_{x_j} F_i({\bf x}) \, [\delta x_j]^{\alpha}  ,\quad i,j=1,...,n.
\ee
From equation (\ref{3-Sta1}), and the property of variation
\be
\delta^{\alpha} D^1_t x_i = D^1_t \delta^{\alpha} x_i,
\ee
we obtain
\be \label{3-fvar3}
D^1_t \, \delta^{\alpha} x_i=
\, _{a_j}^CD^{\alpha}_{x_j} F_i \, [\delta x_j]^{\alpha} ,\quad i=1,...,n.
\ee
Note that in the left hand side of equation (\ref{3-fvar3}),
we have fractional variation of $\delta^{\alpha} x_i$,
and in the right hand side -
the fractional power of variation $[\delta x_i]^{\alpha}$.

Let us consider the fractional variation of the variable $x_i$.
Using equation (\ref{3-fv}), we obtain
\be \label{3-yy}
\delta^{\alpha} x_i=
\, _{a_j}^CD^{\alpha}_{x_j} x_i \, [\delta x_j]^{\alpha}  ,\quad k=1,...,n.
\ee
For the Caputo fractional derivative, we have
\be \label{3-25}
_{a_j}^CD^{\alpha}_{x_j} x_i = \delta_{ij} \ _{a_i}^CD^{\alpha}_{x_i} x_i ,
\ee
where $\delta_{ij}$ is the Kronecker symbol.
Substituting Eq. (\ref{3-25}) into Eq. (\ref{3-yy}),
we can express the fractional power of variations $[\delta x_i]^{\alpha}$
through the fractional variation $\delta^{\alpha} x_i$:
\be \label{3-yy2}
[\delta x_j]^{\alpha} =
\Bigl(\, _{a_j}^CD^{\alpha}_{x_j} x_j \Bigr)^{-1}
\delta^{\alpha} x_j .
\ee
Substitution of (\ref{3-yy2}) into (\ref{3-fvar3}) gives
\be \label{3-fvar2}
D^1_t \delta^{\alpha} x_i = \,
\Bigl(\, _{a_j}^CD^{\alpha}_{x_j} x_j  \Bigr)^{-1} \,
 _{a_j}^CD^{\alpha}_{x_j} F_i ({\bf x}) \, \delta^{\alpha} x_j .
\ee
Here we mean the sum on the repeated index $j$ from 1 to $n$.
Equations (\ref{3-fvar2}) are equations for fractional variations.
Let us denote the fractional variations $\delta^{\alpha} x_k$ by
\be
z_i=\delta^{\alpha} x_i=
 [\delta x_i]^{\alpha} \, _{a_i}^CD^{\alpha}_{x_i} x_i .
\ee

As a result, we obtain the differential equation
for fractional variations
\be \label{3-fxax2}
D^1_t z_i (t) = A_{ij}(\alpha) z_j(t),
\ee
where
\be
A_{ij}(\alpha)= \Bigl( \, _{a_j}^CD^{\alpha}_{x_j} x_j \Bigr)^{-1}
\, _{a_j}^CD^{\alpha}_{x_j} F_i ({\bf x}) . \ee
Using the matrix $Z^t=(z_1,...,z_n)$, and $A_{\alpha}=||A_{ij}(\alpha)||$,
we can rewrite equation (\ref{3-fxax2}) in the matrix form
\be \label{3-Sta2}
D^1_t Z(t) = A_{\alpha} Z(t) . \ee
Equation (\ref{3-Sta2}) is a linear differential equation.
To define the stability with respect to fractional variations,
we consider the characteristic equation
\be
\operatorname{Det} \Bigl( A_{\alpha}-\lambda E \Bigr)=0
\ee
with respect to $\lambda$.
If the real part $Re[\lambda_k]$ of all eigenvalues $\lambda$
for the matrix $A_{\alpha}$ are negative, then the
unperturbed motion is asymptotically stable with
respect to fractional variations.
If the real part $Re[\lambda_k]$ of one of the eigenvalues
$\lambda$ of the matrix $A_{\alpha}$ is positive,
then the unperturbed motion is unstable
with respect to fractional variations.

A system is said to be stable with respect to fractional variations
if for every $\epsilon$, there is a value $\delta_0$ such that:
\be  \|\delta^{\alpha} x(\alpha,t_0)\| < \delta_0 \quad =>
\quad \| \delta^{\alpha} x(\alpha,t)\| < \epsilon \ee
for all $t>t_0$, where $x(\alpha,t)$
describes a state of the system at $t \geqslant t_0$.
The dynamical system is called asymptotically stable
with respect to fractional variations
$\delta^{\alpha} x(t,\alpha)$ if
\be
\lim_{t \to \infty} \|\delta^{\alpha} x(t,\alpha)\| = 0 .
\ee

We note that the notion of stability with respect
to fractional variations (see \cite{FracStab} pages 294-296 and \cite{Book2011})
is wider than the usual Lyapunov or
asymptotic stability \cite{Malkin,Demidovich,Tchetaev}.
Fractional stability includes concept of "integer" stability
as a special case ($\alpha=1$).
A dynamical system, which is unstable with respect to first variation
of states, can be stable with respect to fractional variation.
Therefore fractional derivatives expand our possibility to study
properties of dynamical systems.

As a result, the notion of fractional variations allows us
to define a stability of non-integer order.
Fractional variational derivatives are suggested to describe
the properties of dynamical systems at fractional perturbations.
We formulate stability with respect to motion changes
at fractional changes of variables.
Note that dynamical systems, which are unstable "in sense of Lyapunov",
can be stable with respect to fractional variations.

%%%%%%%%%%%%%%%%%%%%%%%%%%%%%%%%%%%%%%%%%%%%%%%%%%%%%%%%%%%%%%%%%%%%%%%%%%

\end{document}